# Evidence of superconductivity on the border of quasi-2D ferromagnetism in $Ca_2RuO_4$ at high pressure

**Patricia Lebre Alireza (1 and 2), Fumihiko Nakamura (4), Swee Kuan Goh (1), Yoshiteru Maeno (5), Satoru Nakatsuji (6), Yuen Ting Chris Ko (1), Michael Sutherland (1), Stephen Julian (3) and Gilbert George Lonzarich (1)**

(1)Cavendish Laboratory, University of Cambridge, Cambridge UK, (2)Department of Physics and Astronomy, University College London, UK, (3)Department of Physics, University of Toronto, Canada, (4) ADSM, Hiroshima University, Japan, (5)Department of Physics, Kyoto University, Japan, (6)Institute for Solid State Physics, University of Tokyo, Japan.

Email: patricia.alireza@physics.org

## Abstract

The layered perovskite $Ca_2RuO_4$ is a spin-one Mott insulator at ambient pressure and exhibits metallic ferromagnetism at least up to ~ 80 kbar with a maximum Curie temperature of 28 K. Above ~ 90 kbar and up to 140 kbar, the highest pressure reached, the resistivity and ac susceptibility show pronounced downturns below ~ 0.4 K in applied magnetic fields of up to ~10 mT. This indicates that our specimens of $Ca_2RuO_4$ are weakly superconducting on the border of a quasi-2D ferromagnetic state.

## 1. Introduction

Spin-triplet superconductivity in the layered perovskite ruthenate, $Sr_2RuO_4$, is thought to arise via an effective electron-electron pairing interaction that is not due to the mediating role of lattice vibrations alone [1,2]. To throw further light on the possible nature of the pairing interaction we have investigated the temperature-pressure phase diagram of a closely related compound $Ca_2RuO_4$ [3-5], which undergoes a transition from a spin-one Mott insulating state to a metallic magnetic state above ~ 5 kbar [5].

At ambient pressure the layered perovskite structure of $Ca_2RuO_4$ differs from that of $Sr_2RuO_4$ by strong distortions corresponding to compressions, tilts and rotations of the oxygen octahedra [4]. The first of these distortions disappears at room temperature above ~ 5 kbar, where $Ca_2RuO_4$ goes through an insulator to metal transition and begins its transformation into a quasi-2D metallic ferromagnet [5]. The second distortion, or tilt, disappears at ~ 80 kbar, above which the signature for ferromagnetism in the ac magnetic susceptibility appears to be suppressed [6, 7]. Above ~ 90 kbar we find evidence of superconductivity below ~ 0.4 K extending at least up to 140 kbar, the maximum pressure reached in this study.

## 2. Experimental details

Single crystals of $Ca_2RuO_4$ with essentially stoichiometric oxygen content were grown by a floating-zone method with $RuO_2$ self-flux using a commercial infrared furnace (NEC Machinery, SC-M15HD). Our electron-probe microanalyses show no evidence for secondary phases and in particular of Ru inclusions that could give rise to weak parasitic signatures of superconductivity. The samples investigated here came from a batch having residual resistivities of the order of two to three μΩcm in the metallic state above 20 kbar.

Pressures of up to 140 kbar were generated using diamond anvil cells with Daphne oil 7373 as the pressure transmission medium. The sample space was a 350 μm diameter hole drilled in a BeCu gasket pre-indented to a thickness of 70 μm. The culet of the diamond anvils used was 800 μm in diameter. The samples were cleaved into thin rectangular shapes with the shortest sides along the c-axis of the crystal and less than half the pre-indented gasket thickness and the longest sides typically around half the initial diameter of the gasket hole. The pressure in the sample region was measured at room temperature via ruby fluorescence and at low temperatures via the superconducting transition of a tiny Pb sample placed adjacent to the $Ca_2RuO_4$ sample in the ac susceptibility measurements. The pressures measured by these two techniques typically differed by less than 10% and was of the order of the pressure inhomogeneity inferred from the width of the Pb transitions.

The resistivity measurements were carried out via a two-terminal ac technique. Attaching two 12 μm gold wires to the sample via silver epoxy made the current contacts. The total contact resistance was of the order of half an ohm at low temperatures. The voltage contacts were made at the end of the gold wires just outside of the high-pressure space. The ac current was 1.6 μA rms at 73 Hz and the voltage was measured by means of a low-temperature matching transformer (x $10^2$), a low noise preamplifier (x $10^3$) and a phase sensitive detector. The resistivity measurements were carried out with a dilution refrigerator down to 0.04 K. The resistivity was found to be independent of current up to at least ten times the above-given applied current over the entire temperature range investigated.

The ac susceptibility has been measured by the technique described in reference [7] in which a tiny pick-up coil is placed around the sample within the high-pressure region (internal coil method). Corroborating results were also obtained using a small pick-up coil just outside of the high-pressure region (external coil method), together with a high-precision mutual inductance bridge, which could detect changes in the pick-up voltage as low as 0.01 ppm. The external pick-up coil technique was used to investigate the field dependence of the ac susceptibility of the sample as discussed in the next section. All ac susceptibility measurements were carried out with an adiabatic demagnetization refrigerator down to 0.07 K.

## 3. Experimental results and discussions

The results of our studies are shown in figures 1-3. Figure 1 summarizes our measurements of the Curie temperature $T_{Curie}$ and of the superconducting transition temperature $T_{sc}$ versus pressure up to 140 kbar. $T_{Curie}$ and $T_{sc}$ were measured by the ac susceptibility technique based on an internal coil as mentioned in the last section. $T_{Curie}$ was determined from the position of the maximum of the peak in the ac susceptibility versus temperature. The identification of this peak with a ferromagnetic transition has been discussed previously [5, 8]. $T_{sc}$ has been determined from the intersection of the two lines drawn tangent to the ac susceptibility curves in the normal region and in the superconducting region (where the curve drops off precipitously with decreasing temperature).

$T_{Curie}$ rises with increasing pressure to a maximum of 28 K. The signature for $T_{Curie}$ in the ac susceptibility appears to weaken with increasing pressure and could not clearly be identified in our measurements above ~ 80 kbar [8]. We cannot conclude from this, however, that magnetic order is absent above this pressure. The attenuation of the peak may be due to increasing pressure inhomogeneities or intrinsic magnetic inhomogeneities. Evidence for superconductivity in the ac susceptibility was found above about 90 kbar. $T_{sc}$ appears to rise to approximately 0.4 K at 140 kbar, the highest pressure reached.

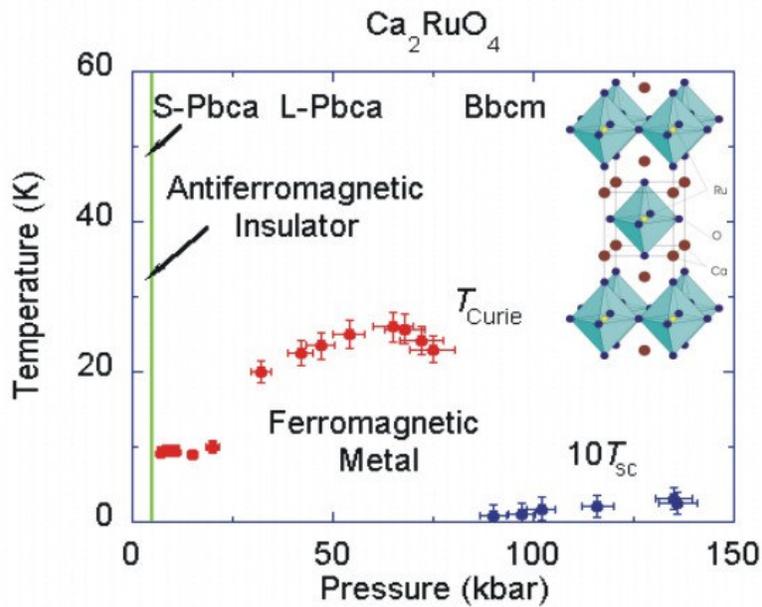

Figure 1. Temperature-pressure phase diagram of $Ca_2RuO_4$. The lattice structure of $Ca_2RuO_4$ is based on the layered perovskite structure shown in the inset, but with distorted oxygen octahedra. At ambient pressure the octahedra are compressed, tilted and rotated (space group S-Pbca) and $Ca_2RuO_4$ is described as a spin-one antiferromagnetic insulator. Above ~ 5 kbar the compression is suppressed (space group L-Pbca) and $Ca_2RuO_4$ transforms gradually into a metallic ferromagnetic state. Above about ~ 80 kbar the tilt is also lost (space group Bbcm) and the ferromagnetic signature ($T_{Curie}$) in the ac susceptibility appears to be suppressed. Above ~ 90 kbar superconducting signatures ($T_{sc}$, magnified by a factor of ten in the figure) are observed. Signatures of superconductivity are corroborated by two ac susceptibility measurement techniques and by measurement of the electrical resistivity as explained in the text.

Figures 2 and 3 show the effect of an applied magnetic field on the temperature dependences of the electrical resistivity and ac susceptibility, respectively, near to the superconducting transition. In all cases the data is collected sweeping up in temperature after cooling in essentially zero external magnetic field (less than a few tenths of a mT).

As shown in figure 2, in the zero-field limit the observed drop in the resistance of the sample below 0.4 K is approximately 5 mΩ. The residual resistance of the sample above this drop can be estimated from the dimensions of the sample (caption of figure 2), the estimate of the basal plane resistivity given in section 2 and a ratio of c-axis to basal plane resistivity assumed to be $10^3$ [3, 5]. In a simple model of sheet resistance for an anisotropic material we estimate the residual resistance to be of the same order of magnitude as the above-given drop in resistance, i.e., as is expected for a superconducting transition.

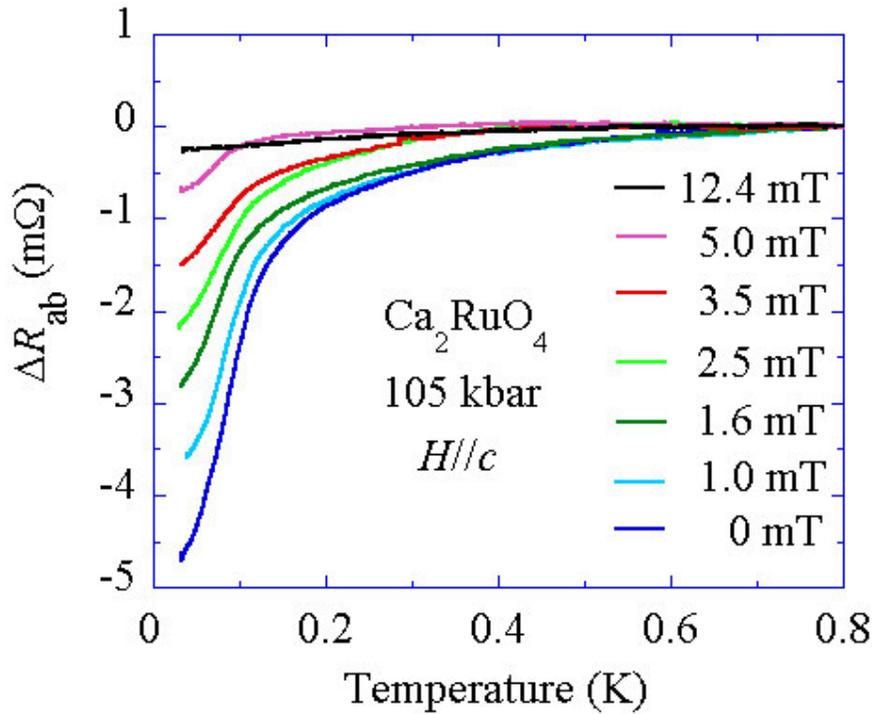

Figure 2. Evidence of superconductivity in the electrical resistivity. The decrease in the resistance of the sample relative to 0.8 K with decreasing temperature at various applied magnetic fields at 105 kbar. The sample was a thin plate of dimensions (180 x 100 x 25) µm, with the normal to the plate parallel to the c-axis of the crystal. Current leads were attached on the plane of the plate with a separation of approximately 100 µm. The field was applied along the c-axis. Further details on the resistivity measurements are given in the text under section 2. The total internal field could differ substantially from the external applied field if magnetic order persists up to 105 kbar.

As shown in figure 3 the observed drop in the pick-up voltage in the ac susceptibility measurement is in excess of 6 nV for our experimental set-up in the zero field limit. We note, however, that the transition is not complete in the temperature range of the measurement (the total drop may be as high as 12 nV) and that an internal field that would partly suppress superconductivity may be present even in the absence of an external field in the pressure range of our experiments. The corresponding change in susceptibility of the sample may be inferred from (i) the sample size, coil geometry and drive current and frequency employed and (ii) by comparing the voltage change with that produced by a reference Pb sample introduced in the same measurement system. Both of these methods suggest that the drop in the pick-up voltage is approximately a quarter of that expected for a full superconducting transition in our $Ca_2RuO_4$ sample.

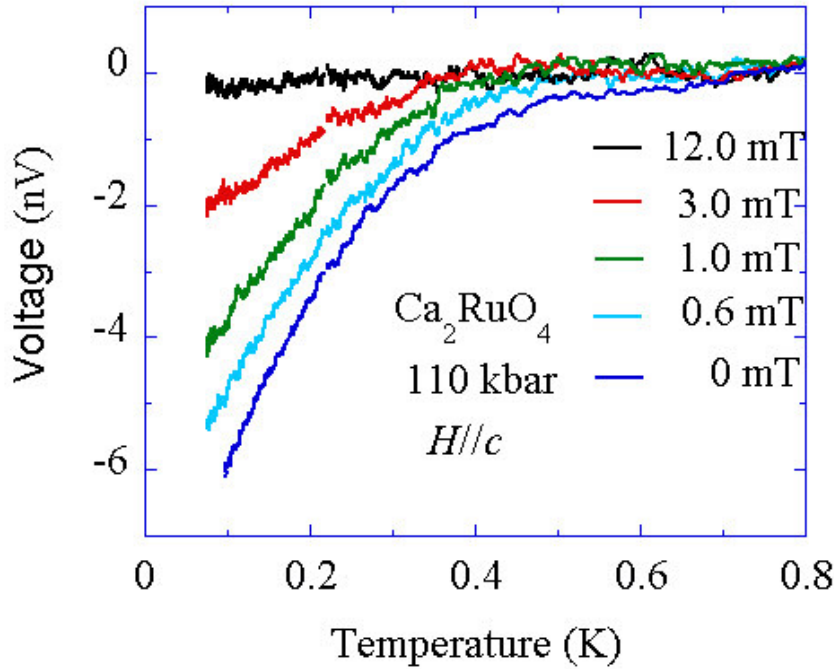

Figure 3. Evidence of superconductivity in the ac susceptibility. The collapse of the ac susceptibility with decreasing temperature relative to 0.8 K at different magnetic fields as seen in the change in the pick-up voltage in a susceptometer at approximately 110 kbar. The sample was in the form of a plate with dimensions (160 x 150 x 25) μm with the normal to the plate along the c-axis of the crystal. The drive coil was made of 200 turns of 30 μm diameter copper wire. The external pick-up coil was made of 140 turns of 15 μm diameter copper wire. Both coils are outside the high-pressure region and wound concentrically around the diamond anvil above the sample. The drive current was 1.6 mA rms and the frequency 2.5 kHz. The pick-up voltage shown corresponds to the component 90° out of phase with the drive current. A field-independent background signal arising mainly from impurities in the BeCu gasket in the diamond anvil cell has been subtracted from the data to emphasize the effect of magnetic field. The ac susceptibility shows a strong magnetic field dependence qualitatively similar to that seen in the resistivity (figure 2). The signature for superconductivity has been seen by two ac susceptibility measurement techniques, as explained in the text, and in over ten specimens at applied pressures in excess of ~ 90 kbar.

As can be seen from figures 2 and 3, superconductivity in $Ca_2RuO_4$ is remarkably fragile and can be largely suppressed with an applied magnetic field of less than 10 mT. Because large changes in the ac susceptibility and resistivity are seen in fields as low as 1 mT, it was necessary to carefully monitor the field in the region of the sample and compensate for residual fields from superconducting magnets and surrounding magnetic materials. The relatively small size of the critical field for superconductivity is due in part to the low $T_{sc}$, but also perhaps because the applied field may only be a part of the total internal field acting on the Cooper pair. An understanding of the low critical field in $Ca_2RuO_4$ awaits studies of the detailed nature of the magnetic state above 80 kbar.

We have presented a temperature-pressure phase diagram of $Ca_2RuO_4$, which suggests that a weakly superconducting state exists on the border of ferromagnetism. This is in contrast to the cuprates in which strong superconductivity appears to be closely connected with antiferromagnetism. It is possible that as in the case of $Sr_2RuO_4$ Cooper pairs in $Ca_2RuO_4$ form in a p-wave spin-triplet state [9-11]. P-wave pairing with low $T_{sc}$ on the border of ferromagnetism, and d-wave pairing with high maximum $T_{sc}$ on the border of antiferromagnetism in layered systems such as the ruthenates and cuprates, respectively, is in keeping with models of superconductivity without phonons based on Cooper-pair formation via the effects of magnetic interactions [see e.g., 11, 12].

**Acknowledgements**

We are grateful for the technical help from Mr Sam Brown who made the pressure cells used in these experiments and Y. Shibata who measured EPMA at the N-BARD, Hiroshima University

We thank Y. Senoo and R. Nakai for their help in the crystal preparation as well as L. Klintberg and E. O'Farrell for their help in running the dilution refrigerator.

This work has been supported by (i) a Grant-in-Aid for Scientific Research on Priority Areas (Grant No. 20029017) from the MEXT of Japan, (ii) the Engineering and Physical Sciences Research Council (EPSRC) of the UK, (iii) the Isaac Newton Trust, (iv) Trinity College, Cambridge, (v) The Croucher Foundation, Hong Kong and (vi) the L'Oreal-UNESCO For Women in Science Fellowship.

**References**

[1] Maeno Y, Hashimoto H, Yoshida K, Nishizaki S, Fujita T, Bednorz JG and Lichtenberg F 1994 Superconductivity in a layered perovskite without copper *Nature* **372** 532-534
[2] Mackenzie AP and Maeno Y 2003 The superconductivity of $Sr_2RuO_4$ and the physics of spin triplet pairing *Rev. Mod. Phys.* **75** 657-712
[3] Nakatsuji S, Ikeda S and Maeno Y 1997 $Ca_2RuO_4$: New Mott insulators of layered ruthenates *J. Phys. Soc. Jpn* **66** 1868-1871
[4] Braden M, Andre G, Nakatsuji S and Maeno Y 1998 Crystal and magnetic structure of Ca2RuO4: Magnetoelastic coupling and the metal-insulator transition *Phys. Rev. B* **58** 847-861
[5] Nakamura F, Goko T, Ito M, Fujita T, Nakatsuji S, Fukazawa H, Maeno Y, Alireza PL, Forsythe D and Julian SR 2002 From Mott insulator to ferromagnetic metal: A pressure study of $Ca_2RuO_4$ *Phys. Rev. B* **65** 220402(R).
[6] Steffens P *et al.* 2005 High-pressure diffraction studies on $Ca_2RuO_4$ *Phys. Rev. B* **72** 094104
[7] Alireza PL and Julian SR 2003 Susceptibility measurements at high pressures using a microcoil system in an anvil cell *Rev. Sci. Instrum.* **74** 4728-4731
[8] Alireza PL, Barakat S, Cumberlidge A-M, Lonzarich GG, Nakamura F and Maeno Y 2007 Developments of susceptibility and magnetization measurements under high hydrostatic pressure *J. Phys. Soc. Jpn* **76 Supplement A** 216-218
[9] Nomura T, and Yamada K 2003 Theory of superconducting mechanism and gap structure of $Sr_2RuO_4$ *Physica C: Superconductivity* **388-389** 495-496
[10] Mazin I and Singh DJ 1999 Competitions in layered ruthenates: Ferromagnetism versus antiferromagnetism and triplet versus singlet pairing *Phys. Rev. Lett.* **82** 4324-4327
[11] Honerkamp C and Rice TM 2003 Cuprates and ruthenates: Similarities and differences *J. Low Temp. Phys.* **131** 61-70
[12] Monthoux P, Pines D and Lonzarich GG 2007 Superconductivity without phonons *Nature* **450** 1177-1183